\documentclass{ws-mpla}

\usepackage{mathtools,amsmath,amssymb,amsfonts,dsfont,mathrsfs}
\usepackage{graphicx}
\usepackage{centernot}
\usepackage{xcolor}
\usepackage{comment}
\usepackage{feynmf}
\usepackage{siunitx}
\usepackage{array}
\usepackage{braket}
\usepackage{xparse}

\DeclareDocumentCommand{\Ag}{ s o }{ \IfBooleanTF{#1}
    { \IfValueTF{#2}{ \boldsymbol{\mathcal{A}}_{(#2)} }{ \boldsymbol{\mathcal{A}} } }
    { \IfValueTF{#2}{            {\mathcal{A}}_{(#2)} }{            {\mathcal{A}} } } }
\DeclareDocumentCommand{\Af}{ s o }{ \IfBooleanTF{#1}
    { \IfValueTF{#2}{ \boldsymbol{A}_{(#2)} }{ \boldsymbol{A} } }
    { \IfValueTF{#2}{            {A}_{(#2)} }{            {A} } } }

\newcommand{\cdf}[1][]{{\boldsymbol{\mathcal{D}}}{#1}}

\newcommand{\df}[1][]{\mathbf{d}{#1}}

\DeclareDocumentCommand{\Fg}{ s o }{ \IfBooleanTF{#1}
    { \IfValueTF{#2}{ \boldsymbol{\mathcal{F}}_{(#2)} }{ \boldsymbol{\mathcal{F}} } }
    { \IfValueTF{#2}{            {\mathcal{F}}_{(#2)} }{            {\mathcal{F}} } } }
\DeclareDocumentCommand{\Ff}{ o }{ \IfValueTF{#1}{ F_{(#1)} }{ F } }

\newcommand{\ga}{\gamma}

\newcommand{\Lag}{\mathscr{L}}

\DeclareDocumentCommand\Te{o o m }{\mathcal{T}{}^{#1}_{#2}(#3)}

\newcommand{\tor}{\mathcal{T}}

\newcommand{\we}{{\scriptstyle\wedge}}

\newcommand{\Bps}{\ensuremath{\bar{\Psi}}}

\newcommand\VIF[1]{\hat{\boldsymbol{e}}^{\hat{#1}}}

\newcommand\hvif[1]{\hat{\boldsymbol{e}}^{{#1}}}

\newcommand\vi[2]{e^{{#1}}_{{#2}}}

\newcommand\vif[1]{\boldsymbol{e}^{{#1}}}

\newcommand\SPIF[2]{\hat{\boldsymbol{\omega}}^{\hat{#1}}{}_{\hat{#2}}}
\newcommand\spi[1]{\omega_{{#1}}}

\newcommand\spif[2]{{\boldsymbol{\omega}}^{{#1}}{}_{{#2}}}

\newcommand\hspi[1]{\hat{\omega}_{{#1}}}

\newcommand{\hRif}[2]{\hat{\boldsymbol{\mathcal{R}}}^{{#1}}{}_{{#2}}}
\newcommand{\Rif}[2]{\boldsymbol{\mathcal{R}}^{{#1}}{}_{{#2}}}

\newcommand{\Tf}[1]{\boldsymbol{\mathcal{T}}^{#1}}

\newcommand{\TF}[1]{\hat{\boldsymbol{\mathcal{T}}}^{\hat{#1}}}

\newcommand{\hcont}[3]{\hat{\mathcal{K}}_{#1}{}^{#2}{}_{#3}}

\newcommand{\CONTF}[2]{\hat{\boldsymbol{\mathcal{K}}}^{\hat{#1}}{}_{\hat{#2}}}

\renewcommand{\(}{\left(}
\renewcommand{\)}{\right)}

\newcommand*{\de}[1]{\mathop{\mathrm{d}#1}\nolimits}

\newcommand\UTFSM{Departamento de F\'\i sica, Universidad T\'{e}cnica Federico Santa Mar\'\i a, \\ Casilla 110-V, Valpara\'\i so, Chile}
\newcommand\CCTVal{Centro Cient\'\i fico Tecnol\'ogico de Valpara\'\i so, \\ Casilla 110-V, Valpara\'\i so, Chile}

\usepackage{url}
\usepackage[super]{cite}
\usepackage[%
  colorlinks=true,
  urlcolor=blue,
  linkcolor=blue,
  citecolor=blue
]{hyperref}
\usepackage{etoolbox}

\hypersetup{%
  pdftitle={Corrections to Higgs width decay into a fermion pair from gravitational torsion},
  pdfauthor={Oscar Castillo-Felisola},
  pdfkeywords={Higgs Decay,} {Beyond Standard Model,} {Torsion,} {Generalised Gravity.},
  pdflang={English}
}

\begin{document}

\markboth{O. Castillo-Felisola, B. D\'iaz S\'aez, F. Rojas, J. Zamora-Sa\'a and A. R. Zerwekh}
{Corrections to Higgs width decay into a fermion pair from gravitational torsion}

\catchline{}{}{}{}{}

\title{CORRECTIONS TO HIGGS WIDTH DECAY INTO A FERMION PAIR FROM GRAVITATIONAL TORSION}

\author{OSCAR {CASTILLO-FELISOLA},\footnote{Corresponding author} \quad BASTI\'AN {D\'IAZ S\'AEZ}, \quad  FELIPE {ROJAS}, \\ JILBERTO {ZAMORA-SA\'A} and ALFONSO R. {ZERWEKH}}

\address{\UTFSM, and \\ \CCTVal. \\ o.castillo.felisola@gmail.com}

\maketitle

\pub{Received (Day Month Year)}{Revised (Day Month Year)}

\begin{abstract}
  We consider the change of the Higgs width decay into a fermion pair  with respect to the standard model, due to the four-fermion contact interaction coming from the existence of gravitational torsion within the context of extra dimension scenarios.
\end{abstract}

\ccode{PACS Nos.: 04.50.Kd,04.90.+e,12.60.-i}

\section{Introduction}\label{intro}

The recent discovery of a bosonic resonance with mass of \SI{125.6}{\GeV} at the LHC~\cite{Aaltonen:2012qt,Aad:2012tfa,Chatrchyan:2012ufa}, has open the possibility of studying in details the electroweak symmetry breaking sector. The current data suggest that the resonance resembles the standard Higgs boson.   Since no deviation from the standard model (SM) has been already observed, it is possible that a New Physics sector may still exist at an energy scale high enough to escape direct detection at the LHC. In such a case, the presence of New Physics can be manifested, although more subtly, through corrections to the properties of the standard particles induced at loop level. For instance, it is expected that  precision measurements of the properties of the Higgs boson at a Higgs factory (for example, the planned International Linear Collider or ILC) could shed light on the nature of an eventual nonstandard sector.

One of the best motivated extensions of the SM is the possible existence of more than three spatial dimensions. In this scenario, it is tempting to consider in the bulk an extended gravitational sector. Indeed, it seems that Einstein's theory of gravity, known as General Relativity (GR), is a low energy effective theory of a more fundamental one due to the lack of a mechanism for quantizing the theory.\footnote{There are several attempts of quantize the gravitational interactions, see for example Refs.~\refcite{Ashtekar:1986yd,Ashtekar:1987gu,Ashtekar:2004eh,DeWitt:1967yk,DeWitt:1967ub,DeWitt:1967uc} (for a historical review see Ref.~\refcite{Rovelli:2000aw}).} In an effort to obtain a more fundamental theory of gravity, several generalizations of GR have been proposed. From these theories, the minimal generalization was proposed by Cartan~\cite{Cartan-Einstein,Cartan1922,Cartan1923,Cartan1924,Cartan1925}, and it is known as Einstein--Cartan theory of gravity (ECT). The main difference between GR and ECT is that the latter does not assume the gravitational connection to be the one of Levi-Civita, and therefore there is an extra component of the connection known as (gravitational) torsion. It is worth to mention that in this minimal frame, the gravitational torsion turns to be a non-dynamical field and can be integrated out of the system.\footnote{Interestingly, for pure gravity the equation of motion for the torsion yields the vanishing torsion condition.}

When the ECT of gravity is coupled with fermionic matter, despite the fact that torsion can still be integrated out, its integration produces a four-fermion contact interaction~\cite{Kibble:1961ba,RevModPhys.48.393,Shapiro:2001rz,SUGRA-book}. In four dimensions the effective four-fermion interaction term has a coupling constant proportional to Newton's gravitational constant, $G_N\sim M_{\text{pl}}^{-2}$, which heavily suppresses the possible phenomenology coming from this term.\footnote{As remarked by L. Fabbri, the most general torsional generalization of Einstein gravity, the effective four-fermion interaction term has a coupling constant proportional to a yet undetermined constant~\cite{Fabbri:2011kq}.} However, in the last decades scenarios with extra dimensions have been proposed as a way to achieve naturalness between the electroweak and the (fundamental) gravitational scales, $M_*$, while the known Planck's mass, $M_{\text{pl}}$, is an enhanced effective gravitational scale~\cite{ADD1,AADD,ADD2,RS1,RS2}.


This new interaction induced by torsion may originates observable effects such as explaining the origin of fermion masses~\cite{Castillo-Felisola:2013jva}, several cosmological problems~\cite{Poplawski:2010jv,Poplawski:2010kb,Poplawski:2011xf,Poplawski:2011wj,Fabbri:2012yg,Vignolo:2014wva}, neutrino oscillation phenomena~\cite{Capozziello:2013dja}, impose limits on extra dimensional model~\cite{Chang:2000yw,Lebedev:2002dp,CCSZ,Castillo-Felisola:2014iia}, and changing one-loop observable~\cite{Castillo-Felisola:2014xba}. A possible  effect of this four-fermion interaction is to modify , through loop effects, the decay width of the Higgs boson into a pair of fermions. This deviation from standard predictions could be observed in principle, by mean of precision measurements performed in a future lepton collider. The  aim of this work is to estimate the size of this effect depending on the extra dimension scale of gravity. To this end, a brief review of the theoretical setup is presented in Sec.~\ref{CEG}. Then, in Sec.~\ref{1loop} we show the one-loop corrections to the Higgs decay into a lepton pair, due to the effective four-fermion interaction. Finally, in Sec.~\ref{phenom} we analyze the phenomenological constraints to the parameter of our model imposed by the experimental data.

\section{Effective interaction through gravitational torsion}\label{CEG}

The standard formalism used in the GR, where the physical field is the metric, is know as second order formalism --- due to the fact that the equations of motion are of second order ---. However, in this section we shall deal with the first order formalism,\footnote{Additionally, we make extensive use of the formalism of differential forms~\cite{Cartan-calc,Zanelli:2005sa}.} which accomplish the same goal through the splitting the equations of motion into first order differential equations  by introducing two independent fields, known as vielbeins \mbox{($\vif{a} = \vi{a}{\mu}\de{x}^\mu$)} and spin connection \mbox{($\spif{ab}{} = \spi{\mu}{}^{ab}\de{x}^\mu$).} These fields define the torsion and curvature two-forms via the Cartan structure equations
\begin{equation}
  \df[\vif{a}] + \spif{a}{b} \we \vif{b} = \Tf{a} \quad \text{and} \quad \df[\spif{ab}{}] + \spif{a}{c} \we \spif{b}{c} = \Rif{ab}{}.
\end{equation}
We shall use the notation in Refs.~\refcite{Castillo-Felisola:2013jva,Castillo-Felisola:2014iia,Castillo-Felisola:2014xba}, where hatted quantities refer to higher dimensional objects and/or indices, and $\ga^{*}$ is the four-dimensional chiral matrix.

We start from the $D$-dimensional action which includes the ECT of gravity and Dirac fields interacting with the gravitational field,\footnote{We assume that fermion masses are developed through the Higgs mechanism, so the is no need for considering nontrivial fundamental mass terms.}
\begin{equation}
  S = \frac{1}{2\kappa^2}\int\frac{\epsilon_{\hat{a}_1\cdots \hat{a}_D}}{(D-2)!}\hRif{\hat{a}_1 \hat{a}_2}{} \we \hvif{\hat{a}_3} \we \cdots \we \hvif{\hat{a}_D} - \int \frac{\epsilon_{\hat{a}_1\cdots \hat{a}_D}}{(D-1)!} \Bps \hvif{\hat{a}_1} \we \cdots \we \hvif{\hat{a}_{D-1}}\ga^{\hat{a}_D} \hat{\cdf} \, \Psi
  \label{action}
\end{equation}
where $\hat{\cdf}$ is the spinorial covariant derivative in a curved spacetime, defined by\footnote{Hereon, multi-index gamma matrices represent the totally anti-symmetric product of gammas.}
\begin{equation*}
  \hat{\cdf} \Psi = \df[\Psi] + \frac{1}{4} (\hspi{\mu})^{\hat{a} \hat{b}} \ga_{\hat{a} \hat{b}} \Psi.
\end{equation*}
The equation of motion for the spin connection in Eq.~\eqref{action} yield an algebraic equation for the components of the torsion, $\TF{a} = \tfrac{1}{2} \hat{\tor}{}_{\hat{b}}{}^{\hat{a}}{}_{\hat{c}} \, \VIF{b} \we \VIF{c}$, 
\begin{equation}
  \frac{1}{2}\(\hat{\tor}_{\hat{b} \hat{c} \hat{a} } + \hat{\tor}_{\hat{b} \hat{a} \hat{c} } + \hat{\tor}_{\hat{a} \hat{b} \hat{c} }\) = -\frac{\kappa^2}{4} \bar{\Psi}\ga_{\hat{a} \hat{b} \hat{c}}\Psi,
  \label{tor-eom}
\end{equation}
where the torsional construction in the LHS is known as the contorsion, $\hcont{\hat{a} \hat{b} \hat{c} }{}{}$, and additionally from the RHS the only not trivial contribution of the contorsion is the totally antisymmetric part. Moreover, the contorsion appears as a tensor which relates the ``affine'' spin connection ($\SPIF{a}{b}$) with the torsion free one ($\hat{\overline{\boldsymbol{\omega}}}^{\hat{a}}{}_{\hat{b}}$), i.e., \mbox{$\SPIF{a}{b} = \hat{\overline{\boldsymbol{\omega}}}^{\hat{a}}{}_{\hat{b}}+\CONTF{a}{b}.$}  

Since the  Eq.~\eqref{tor-eom} is algebraic, it can be substituted back into the original action. The new action, expressed in terms of torsion-free quantities includes GR coupled with Dirac fields, plus an  induced four-fermion contact interaction of the form
\begin{equation}
  \Lag_{\text{4FI}} = \frac{\kappa^2}{32} \( \bar{\Psi}\ga_{\hat{a} \hat{b} \hat{c}}\Psi \)  \( \bar{\Psi}\ga^{\hat{a} \hat{b} \hat{c}}\Psi \),
  \label{Lag4FI}
\end{equation}
which in four dimensions --- where $\kappa^2 = {1}/{M_{\text{Pl}}^2}$ --- is suppressed by the Planck mass as anticipated. Since the value of the Planck mass is several orders of magnitude higher than any order scale in the SM of particle physics, this effective interaction is negligible for any phenomenological effect.

Nevertheless, in the last decades some higher dimensional scenarios have been proposed such that there is a fundamental scale of gravity ($M_*$) which is enhanced in the four-dimensional effective theory up to the Planck mass~\cite{ADD1,AADD,ADD2,RS1,RS2}. In these scenarios, the gravitational scale can be so low as a few \si{\TeV}, providing a solution to the huge difference between the SM scale and gravitational scale, known as the hierarchy problem. Additionally, the change in the gravitational scale have repercussions in the suppression of the four-fermion interaction, because $\kappa$ in Eq.~\eqref{Lag4FI} is replaced by $\kappa_*$.

If we restrict ourselves to consider a single extra dimension, the Clifford algebra in five dimensions can be decomposed in terms of four-dimensional one as follows,
\begin{align}
  \big(\ga_{\hat{a} \hat{b} \hat{c}}\big) \big(\ga^{\hat{a} \hat{b} \hat{c}}\big) = \big(\ga_{a b c}\big)\big(\ga^{a b c}\big) + 3 \big(\ga_{a b *}\big)\big(\ga^{a b *}\big).
\end{align}
Therefore, the interaction in Eq.~\eqref{Lag4FI} give rise to axial--axial and tensor-axial--tensor-axial interactions~\cite{Castillo-Felisola:2013jva}
\begin{equation}
  \begin{split}
    \Lag_{\text{eff}}
    & = \frac{3 \kappa_*^2}{16} \( \bar{\Psi} \ga_{a}\ga^* \Psi \) \( \bar{\Psi} \ga^{a}\ga^* \Psi \) \\
    & \quad + \frac{3 \kappa_*^2}{32} \( \bar{\Psi} \ga_{a b}\ga^* \Psi \) \( \bar{\Psi} \ga^{a b}\ga^* \Psi \).
  \end{split}
  \label{Eff4fi}
\end{equation}

\section{One-loop calculation of $H \to \ell \bar{\ell}$}\label{1loop}

\begin{figure}[hb]
  \begin{center}
    \includegraphics[width=.5\textwidth]{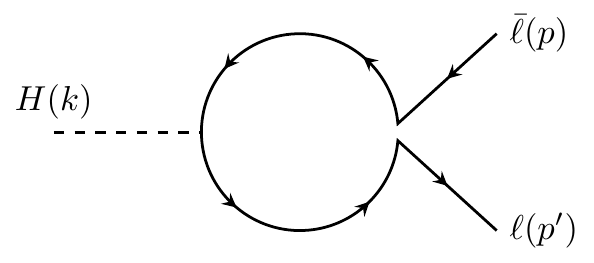}
  \end{center}
  \caption{Higgs to lepton pair through the four-fermion \mbox{``s-channel''.}}
  \label{fig:s}
\end{figure}

\begin{figure}[hbt]
  \begin{center}
    \includegraphics[width=.5\textwidth]{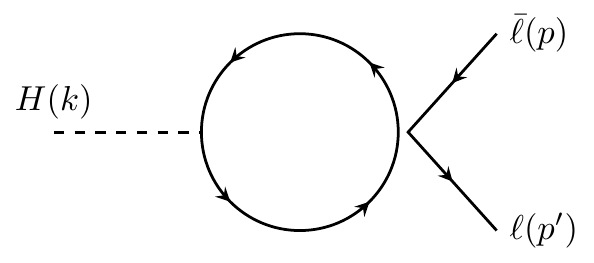}
  \end{center}
  \caption{Higgs to lepton pair through the four-fermion \mbox{``t-channel''.}}
  \label{fig:t}
\end{figure}

In this section we shall consider the contribution of the effective four-fermion interaction in Eq.~\eqref{Eff4fi}, to the decay of the Higgs boson into a lepton pair, which is a one-loop process. To this end, we split the effective interaction into a current--current interaction (inspired by Ref.~\refcite{GonzalezGarcia:1998ay})
\begin{equation}
  \Lag_{\text{eff}} = \frac{3 \kappa_*^2}{16} \big( J_{a}^* \big) \big( J^{a*} \big) + \frac{3 \kappa_*^2}{32} \big( J_{ab}^* \big) \big( J^{ab*} \big),
  \label{new4fi}
\end{equation}
which generate two different contributions to the $H \to \ell \bar{\ell}$ process, which will be called \emph{s-channel} (see Fig.~\ref{fig:s}) and \emph{t-channel} (see Fig.~\ref{fig:t}) respectively.

A (psudo)scalar field decays into a lepton pair through a current of the form
\begin{equation}
  J = \bar{u}_\ell(\vec{p}) \( S + \imath P \ga^* \) v_\ell(\vec{p}'),
  \label{Scur}
\end{equation}
where $S$ and $P$ are the scalar and pseudo-scalar form factors, and $\ga^*$ is the chiral matrix in four dimensions. Accordingly to the current in Eq.~\eqref{Scur}, the width decay of a (pseudo)scalar particle into a fermion pair is given by 
\begin{equation}
  \Gamma(H \to \ell \bar{\ell}) = \frac{g^2}{32} \frac{M_H m_\ell^2}{M_W^2}  \sqrt{1 - \frac{4 m_\ell^2}{M_H^2} }  S^2 \( 1 + \frac{P^2}{S^2} - \frac{4 m_\ell^2}{M_H^2} \).
\end{equation}
Notice that for the Higgs boson $S^{\text{tree}} = 1$ and $P^{\text{tree}} = 0$.

We used the Mathematica package ``FeynCalc''~\cite{Kublbeck:1992mt} to calculate the correction to the Higgs width decay. Although we consider the general current presented in Eq.~\eqref{Scur}, the result shows that no pseudo-scalar form factor is generated, i.e., $P = 0$ to one-loop. Additionally, due to the Lorentz structure of the fermionic currents composing the four-fermion contact interaction, the ``t-channel'' Feynman diagram (Fig.~\ref{fig:t}) does not contribute to the width decay of the Higgs boson.

In order to obtain a numerical result, we assume that the fundamental scale of gravity $M_*$ is of the order of the \emph{new physics} scale $\Lambda$. Therefore, our results, despite of coming from extra dimensions, do not depend explicitly of the number of extra dimensions. We parametrize the corrections to the width decay as a function of this scale of new physics.

In our model, the correction to the scalar form factor induced by the four-fermion interaction is
\begin{equation}
  \delta S = - \frac{15}{64} \frac{1}{\Lambda^2} m_f \( m_H^2 - 2 m_f^2 \) \log\( \frac{\Lambda^2}{m_H^2} \).
\end{equation}
This results into a variation of the width decay  of the form
\begin{equation}
  \delta \Gamma_{\text{4FI}} = -\frac{15}{16384} \frac{g^2 m_H m_f^2}{\pi m_W^2 \Lambda^2} \( m_H^2 - 2 m_f^2 \) \( 1 - \frac{4m_f^2}{m_H^2} \)^{3/2} \log\( \frac{\Lambda^2}{m_H^2} \) .
\end{equation}
The above formula gets modified by a colour factor ($N_c = 3$) when we consider decay into quarks.

\section{Phenomenological implications}\label{phenom}

As mentioned, only the ``s-channel'' Feynman diagram contribute to the variation of the width decay of the Higgs into fermion pairs. Since the torsion induced four-fermion interaction comes from the kinetic term, although the dimensional reduction induces a Kaluza--Klein tower in the effective particle spectrum, indisputably the fermion around the loop has the same flavour as the outgoing particles. Therefore, none of the particles in the Kaluza--Klein tower enter in the analysis.

\begin{figure}[hbt]
  \begin{center}
    \includegraphics[width=.75\textwidth]{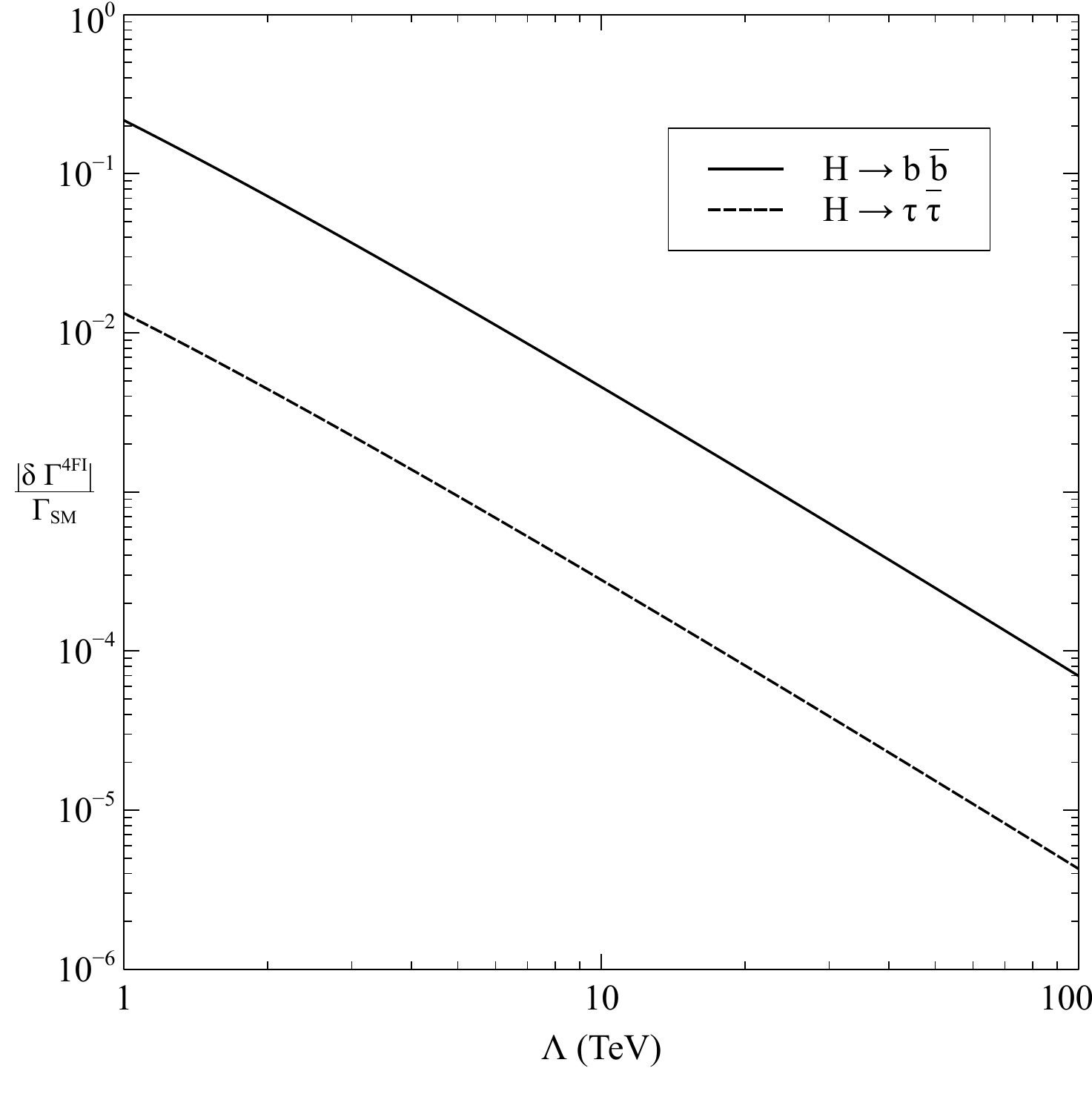} 
  \end{center}
  \caption{Variation to the partial width decay of the Higgs into tau lepton and bottom quark pairs, normalized by the total width decay.}
  \label{fig:Hdt}
\end{figure}

We focus on decays into $\tau^+ \tau^-$ and $b \bar{b}$ pairs, because these are the main fermionic decay modes. In order to estimate the size of the effect we compare it with the total width decay of the Higgs predicted by the SM. The results are shown in Fig.~\ref{fig:Hdt}.

From the plot (Fig.~\ref{fig:Hdt}), it is clear that the variation on the Higgs width decay is one order of magnitude enhanced in the case of the bottom quark pair, in comparison with the decay into $\tau$ pairs. For fundamental gravitational scales as low as \SI{1}{\TeV}, the correction induced by the torsion-induced interaction is of about \num{11}\% for the decay $H \to b\bar{b}$, while for the process $H \to \tau^+ \tau^-$ it decreases to \num{1}\%. It can be expected that effects of these orders could be observed at the Large Hadron Collider (LHC) within a few years, when the luminosity will be improved and the statistics is enough for accomplishing the required precision.

However, recent analysis on the constraints imposed by the torsion induced four-fermion interaction on the $Z$ boson decay (see Refs.~\refcite{Lebedev:2002dp,Castillo-Felisola:2014xba}), the strongest limit is $\Lambda \simeq \SI{30}{\TeV}$. Given this stringent limit, the correction to the width decay of the Higgs drops to approximately \num{.1}\% and \num{.05}\% for bottom and tau pairs respectively. Such limits are unlike to be measure in current experiments, but could be reached at future Higgs factories, such as the International Linear Collider (ILC) or the Compact Linear Collider (CLIC), which could provide a deeper insight of the core process because the electron-positron annihilation is clearer, due to the reduced background.

\section*{Acknowledgements}

B. D.~S. thanks to the Universidad Austral de Chile for the hospitality during the completion of this work. F. R. thanks to the University of Southampton (UK) for the hospitality during the completion of this work.

This work was partially supported by USM grant No. 11.15.77, CONICYT (Chile) under project No. 79140040 and FONDECYT (Chile) projects No. 1120346 and FSM-1204.


\end{document}